# LOW POWER DUAL EDGE - TRIGGERED STATIC D FLIP-FLOP

Anurag[#1], Gurmohan Singh[#2], V. Sulochana[#3]

[#] Centre for Development of Advanced Computing, Mohali, India

[1]anuragece09@gmail.com
[2]gurmohan@cdac.in
[3]vemus@cdac.in

## ABSTRACT

*This paper enumerates new architecture of low power dual-edge triggered Flip-Flop (DETFF) designed at 180nm CMOS technology. In DETFF same data throughput can be achieved with half of the clock frequency as compared to single edge triggered Flip-Flop (SETFF). In this paper conventional and proposed DETFF are presented and compared at same simulation conditions. The post layout experimental results comparison shows that the average power dissipation is improved by 48.17%, 41.29% and 36.84% when compared with SCDFF, DEPFF and SEDNIFF respectively and improvement in PDP is 42.44%, 33.88% and 24.69% as compared to SCDFF, DEPFF and SEDNIFF respectively. Therefore the proposed DETFF design is suitable for low power and small area applications.*

## KEYWORDS

*Dual-Edge Triggered, Flip-Flop, High Speed, Low Power, Static D Flip-Flop*

## 1. INTRODUCTION

The latest advancement in computing technology has set a goal of high performance with low power consumption for VLSI designer [1]. Flip-Flops are important timing elements in digital circuits which have a great impact on circuit power consumption and speed. The performance of the Flip-Flop is an important element to determine the performance of the whole synchronous circuit, particularly in deeper pipelined design [2]. For improving the performance one innovating approach is to increase the clock frequency. However, using high clock frequency has many disadvantages. Power consumption of the clock system increases dramatically and clock uncertainties take significant part of the clock cycle at high frequencies. Moreover the non-ideal clock distribution results in degradation of the clock waveform, power supply noise and cross-talk. About 30%-70% of the total power in the system is dissipated due to clocking network, and the Flip-Flops [3]. An alternative clocking approach is based on the use of storage elements which are capable of capturing data on both rising and falling edges of the clock. Such storage elements are termed as Dual-Edge Triggered Flip-Flops (DETFFs). In this scenario, same data throughput can be achieved at half of the clock frequency as compared to single edge triggered Flip-Flops [4]. In other words we can say that double edge clocking can be used to save half of the power in the clock distribution network. The average power in a digital CMOS circuits is given by the following equation:

$$P_{avg} = p_t(C_L V * V_{dd} * f_{clk}) + I_{sc} * V_{dd} + I_{leakage} * V_{dd} \qquad (1)$$

The above equation represents the three major sources of power dissipation in CMOS VLSI circuits. The first term represents the dynamic or switching power dissipation. The second term indicates the direct path short circuit power dissipation. The third term reflects leakage power. All these three types of powers are highly dependent on supply voltage. In majority of the cases, the voltage swing V is the same as the supply voltage $V_{dd}$. Dynamic power is proportional to the square of the supply voltage, contributes highest power consumption among the three. Therefore reducing the supply voltage is the most effective way to reduce power consumption of the design. However it decreases the speed of the designed circuit. Reduction in clock frequency is another alternative to reduce the dynamic power. Double edge clocking approach is adapted in this paper to reduce the clock frequency. In this approach same data throughput can be achieved with half of the clock frequency as compared to SETFF.

The paper is organized as follows-

Section 2 explains the conventional DETFF circuits and a new proposed architecture of DETFF. Section 3 contains the nominal simulation conditions, along with analysis and optimization performed, during simulation. Section 4 contains performance results for proposed design. These results are compared with conventional designs in terms of delay, power, PDP and area and Section 5 ends with conclusion.

## 2. FLIP FLOP STRUCTURES

In some of the designs DETFF approach is preferred to reduce power dissipation [5]. Unlike SETFF, data is captured by both edges of the clock. Implementation of DETFF is shown in the block diagram in Fig 1.

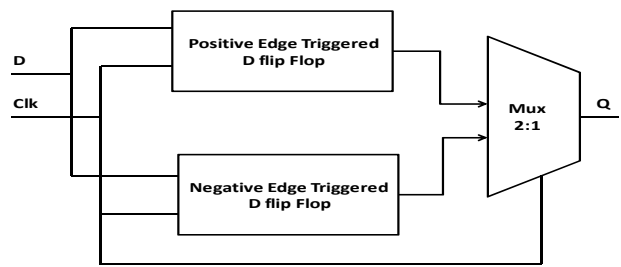

Fig 1 - Dual-edge-triggered Flip-Flops

Both positive and negative edges are used to sample the D input at alternate clock edges, and the appropriate sample is selected for the Q output by a clocked multiplexer (MUX).

### 2.1. Conventional Dual-Edge Triggered Flip-Flops

M.W.Phyu et.al, proposed a static output-controlled discharge Flip–Flop (SCDFF) [6]. SCDFF is implemented by Cross-coupled inverters to keep the data at the output Q. However, race problem is there in the cross-coupled inverters, which not only degrades the speed of charging and discharging, but also causes short-circuit power dissipation. The duration of the race current will be prolonged if the output load capacitance is large, which may distort the desired output signal and also increases the dynamic power dissipation.

Yan-yun Dai et.al, proposed a dynamic explicit-pulsed double-edge triggered Flip–Flop (DEPFF) [7], Which is a pulse-triggered Flip–Flop. To precharge the internal node an always ON PMOS transistor is used in DEPFF pulse generator circuit, but it results in a short-circuit current in the scenario when the discharge path is also ON.

Xue-Xiang Wu et.al, proposed a static explicit-pulsed dual edge triggered Flip-Flop with latch node built-in (SEDNIFF) [8]. In SEDNIFF, a pulse generator circuit is used to generate narrow pulses at both rising and falling edges of the clock as shown in Fig 2.

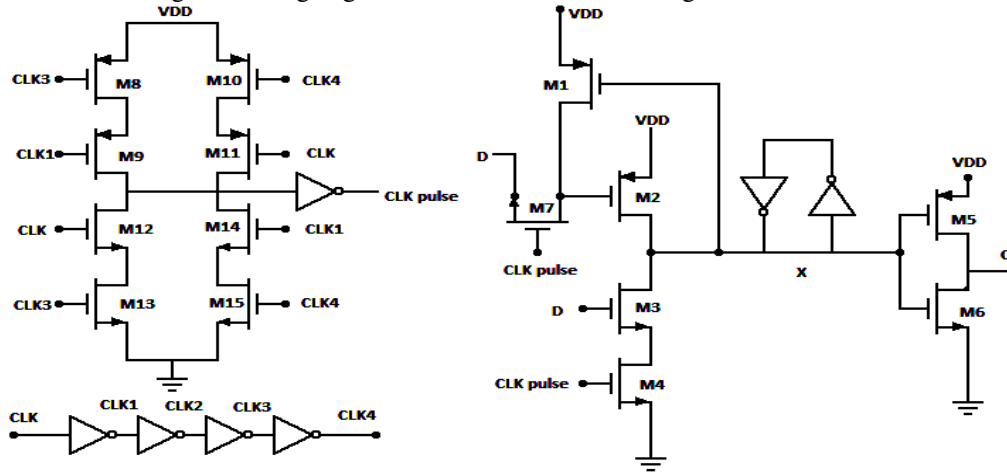

Fig 2- SEDNIFF circuit

In SEDNIFF the charge paths, from VDD to CLK-pulse are OFF when the discharge paths, from CLK-pulse to ground are ON and vice versa, which leads to reduction in short-circuit power dissipation. However because of large number of clocked transistors present in the clock generator circuit the overall power dissipation of the design is more.

## 2.2. Proposed Dual-Edge Triggered Flip-Flop

In the proposed DETFF, positive latch and negative latch are connected in parallel as shown in Fig 3. These latches are designed using one transmission gate and two inverters connected back to back and the output of both the latches are connected to 2:1Mux as input. Mux is designed using one PMOS and one NMOS connected in series and gates are connected together and derived by the inverted CLK. Output of Mux is connected to the inverter for strengthening the output. Back to back connected inverters hold the data when transmission gate is *OFF* and at the same time Mux sends the latched data to the inverter to get the correct D at the output.

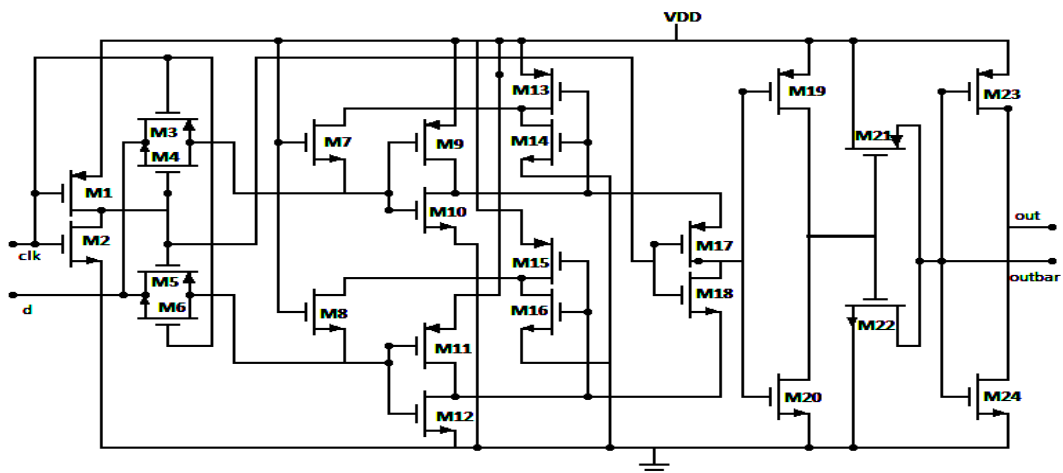

Fig 3 - Proposed DETFF Circuit

The proposed DETFF works as follows. When the CLK is low M3, M4 and M18 are ON and M5, M6 and M17 are OFF. Hence data hold by negative latch is transparent to Q. When CLK is high M5, M6 and M17 are ON and M3, M4 and M18 are OFF. If input D remains the same, Q also remains unchanged. On the other hand, if D is changed before the CLK then D will be hold by positive latch and the same value will be send to the output when CLK changes from Low to high and similarly for the transition of CLK from high to low.

## 3. SIMULATIONS

For a fair comparison each circuit is simulated at the layout level on same simulation parameters. The layout of proposed DETFF shown in Fig 4 is designed using cadence virtuoso layout editor. Simulation parameters used for comparison are shown in Table I.

Table I
CMOS Simulation Parameters

| Technology | 180 nm |
|---|---|
| Min. Gate Width: | 600 nm |
| Max. Gate Width: | 1200 nm |
| MOSFET Model: | BSIM 3v3 |
| Nominal Conditions: | $V_{dd}$ = 1.8V, T=27 $^0$C |
| Duty Cycle: | 50 % |
| Nominal Clock Frequency: | 125MHz |

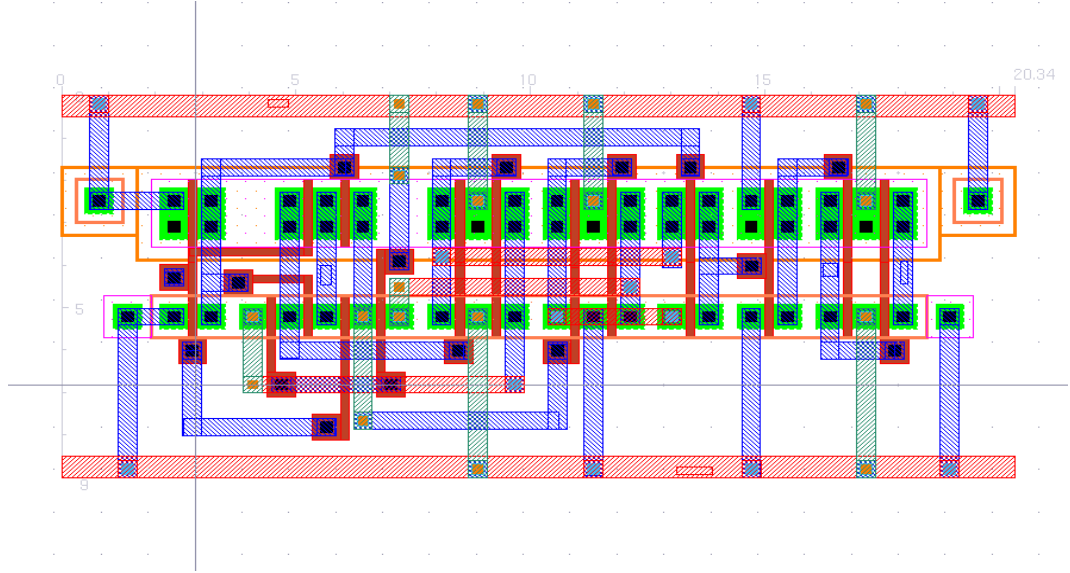

Fig 4- Layout of proposed DETFF

All simulations are performed on Cadence Spectre circuit simulator using BSIM3v3 models at 180nm technology node. The Post-layout simulation waveforms for proposed DETFF are shown in Fig 5. The post-layout simulation results were captured for 125MHz clock frequency and 21fF load at the output. The input data set D (111101011001000) is applied with a pulse width of 7.5ns at a supply voltage of 1.8V for 120ns duration.

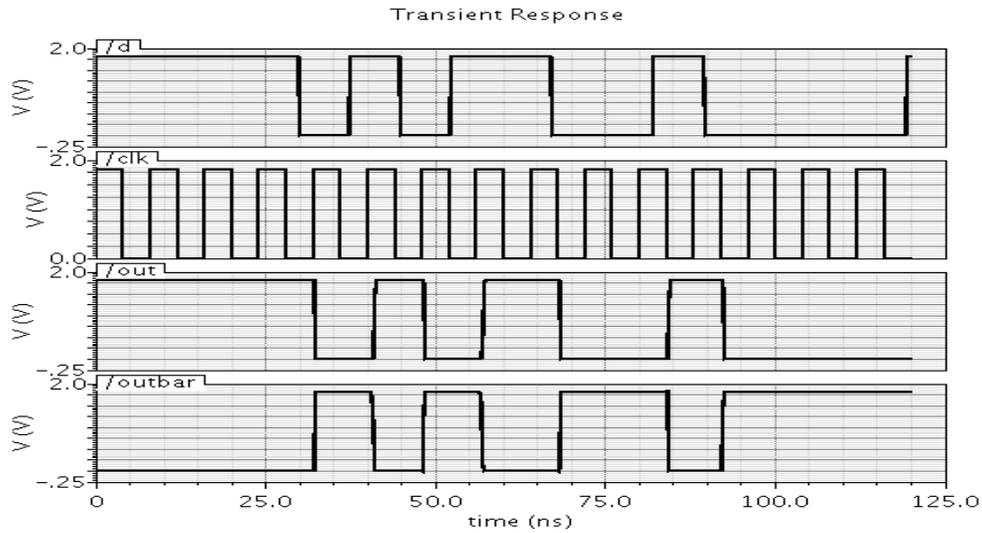

Fig 5 - Output waveform for proposed DETFF

### 3.1. Analysis and Optimization

Flip-Flops designs can be compared at different parameters like delay, average power dissipation, power delay product (PDP) etc. Generally, for portable systems in which the battery life is of the primary concern PDP-based comparison is appropriate [8]. However, there is always a trade off between propagation delay and power dissipation of a circuit. If we optimize a circuit for power its delay increases and vice-versa. We designed the circuit to achieve minimum power-delay product (PDP). In addition to this delay, power consumption and area of the Flip-Flop designs are also compared.

## 4. EXPERIMENTAL RESULTS COMPARISON

The proposed DETFF is designed and compared with several conventional Flip-Flops. Each Flip-Flop is optimized for power delay product. The proposed DETFF is having lesser number of clocked transistors than the other discussed DET FFs. Simulation results for power, delay, PDP and area at nominal conditions for the Flip-Flops are summarized in Table II.

Table II
Results Comparison

| Design Name | Layout Area (μm²) | Number of Transistors used | Minimum Clk-to-Q Delay (ps) | Total avg. Power (μW) | PDP (fj) |
|---|---|---|---|---|---|
| SCDFF | 682.2 | 29 | 234.5 | 41.97 | 9.80 |
| DEPFF | 578.3 | 29 | 230.2 | 37.05 | 8.53 |
| SEDNIFF | 497.7 | 29 | 217.7 | 34.44 | 7.49 |
| Proposed DETFF | 183.06 | 24 | 259.6 | 21.75 | 5.64 |

Post-layout simulation results show that the proposed DET flip-flop has 21.75µW average power dissipation that shows an improvement of 48.17%, 41.29% and 36.84% when compared with SCDFF, DEPFF and SEDNIFF respectively. On the other hand clock to Q delay is increased by 10.7%, 12.77% and 19.24% respectively. However proposed DETFF has an improvement of 42.44%, 33.88% and 24.69% in terms of power delay product (PDP) as compared to SCDFF, DEPFF and SEDNIFF respectively. The proposed design also has an improvement of 73.16%, 68.34% and 63.21% in terms of layout area as compared to SCDFF, DEPFF and SEDNIFF respectively.

# 5. CONCLUSION

In this paper, we proposed a low power, small area DETFF design which is static in nature. The proposed DETFF has lesser number of clocked transistors with respect to other DETFF. The post layout experimental simulation results shows that proposed DETFF offers improvement in power dissipation, PDP and area. Therefore the proposed DETFF is very well suited for low power and small area applications.

# Authors

Anurag is pursuing Masters of Technology at C-DAC Mohali in VLSI Design. He has obtained his Bachelor of Engineering degree in Electronics & Communication Engineering from Kumaun University, Nainital in 2009. His research interests include low power Digital & Analog VLSI Design. His email-id is anuragece09@gmail.com

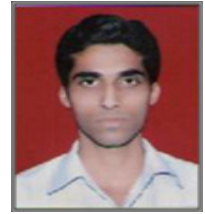

Gurmohan Singh has obtained his Bachelor of Technology degree in Electronics & Communication Engineering from Giani Zail Singh College of Engineering & Technology, Bathinda and Master of Technology degree in Mircoelectronics from Panjab University, Chandigarh in 2001 and 2005 respectively. He is working as a Senior Engineer in Digital Electronics & Comm. Division at C-DAC, Mohali. He is involved in many technological research areas in the field of VLSI Design. He has more than 7 years of experience. He has also worked in Bharat Sanchar Nigam Limited for 3 years. He was invoved in installation, configuration & Testing of latest communication equipements like DWDM, GPON OLTEs & LAN Switches and MADMs etc. His major research interests are analog integrated circuits design, low power CMOS circuit design techniques and VLSI Testing. His email-id is gurmohan@cdac.in

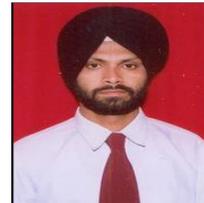

Vemu Sulochana has obtained her Bachelor of Technology degree from JNTU Kakinada and Master of Technology degree from NIT, Hamirpur in 2004 and 2009 respectively. In 2011, she joined C-DAC, Mohali to conduct innovative research in the area of VLSI design, where she is now a Project Engineer - II. Her research is concerned with low power VLSI design, Design of high speed VLSI interconnects. She is conducting research in IC interconnect characterization, modeling  and simulation for the high speed VLSl circuit design.Her email-id is vemus@cdac.in

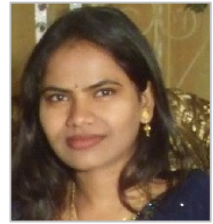